\documentclass[twocolumn,showpacs,preprintnumbers,amsmath,amssymb]{revtex4}

\usepackage{graphicx}
\usepackage{dcolumn}
\usepackage{bm}
\usepackage{float}
\usepackage{natbib}
\usepackage{amssymb}

\begin{document}

\title{Pure single photons from a trapped atom source }
\author{D. B. Higginbottom$^{1,2}$}
\email{daniel.higginbottom@anu.edu.au}
\author{L. Slodi\v{c}ka$^{3}$}
\author{G. Araneda$^{2}$}
\author{L. Lachman$^3$}
\author{R. Filip$^3$}
\author{M. Hennrich$^2$}
\author{R. Blatt$^{2,4}$}
\affiliation{$^1$Australian National University, Canberra ACT 0200, Australia}
\affiliation{$^2$Institut f\"{u}r Experimentalphysik, Universit\"{a}t Innsbruck, Technikerstr. 25, 6020 Innsbruck, Austria}
\affiliation{$^3$Department of Optics Palack\'{y} University, 17. Listopadu 12, 77146 Olomouc, Czech Republic}
\affiliation{$^4$Institut f\"{u}r Quantenoptik und Quanteninformation, \"{O}sterreichische Akademie der Wissenschaften, Technikerstra\ss e 21a, 6020 Innsbruck, Austria}

\begin{abstract}

Single atoms or atom-like emitters are the purest source of
on-demand single photons, they are intrinsically incapable of
multi-photon emission. To demonstrate this degree of purity we
have realised a tunable, on-demand source of single photons using
a single ion trapped at the common focus of high numerical
aperture lenses. Our trapped-ion source produces single-photon
pulses at a rate of 200 kHz with g$^2(0) = (1.9 \pm 0.2) \times
10^{-3}$, without any background subtraction. The
corresponding residual background is accounted for exclusively by
detector dark counts. 
We further characterize the performance of our source by
measuring the violation of a non-Gaussian state witness and show
that its output corresponds to ideal attenuated single photons.
Combined with current efforts to enhance collection efficiency from single emitters, our results suggest that single trapped ions are not only ideal stationary qubits for quantum information processing, but
promising sources of light for scalable optical quantum networks.


\end{abstract}

\pacs{42.50.-p, 42.50.Ar}

\maketitle

Light fields of a definite, single excitation are an elementary tool for quantum information and a key feature in schemes for encoding, manipulating and communicating quantum information. A number of protocols for unconditionally secure communication harness the indivisibility of single photon states to distribute an unread cryptographic key (QKD) \cite{Bennett1984, Ekert1991}. Over long distances QKD requires remote entanglement distributed over optical repeater networks, and these in turn require further quantum optical resources. Moreover, to scale effectively with distance single-photon repeater nodes must be capable of producing single photons on demand with a critically low multi-photon component \cite{Dutton2014}. The search for high purity single-photon sources is further motivated by their proposed application in systems for performing quantum simulations \cite{Georgescu2014} and computing quantum algorithms without an efficient classical equivalent \cite{Knill2001a, Kok2007}. Such optical quantum computers demand many identical sources for delivering strictly indistinguishable single photons on demand.

An ideal single photon source emits a single photon (and never more than one photon) on demand with 100\% probability, at an arbitrarily high rate, and indistinguishable both from previous emissions and from parallel sources \cite{Eisaman2011}. Existing sources differ substantially from this ideal. The workhorse of current quantum information experiments \cite{Soujaeff2007} is spontaneous parametric down conversion (SPDC) which probabilistically produces correlated pairs of photons across two modes, one of which can be used to herald a single photon in the other. To operate `on demand' probabilistic sources like SPDC must be combined with a suitable quantum memory, however even as a probabilistic source SPDC has limitations. Unless efficient photon number resolving detectors are available to discriminate single-pair from multi-pair events at the herald, SPDC sources produce fields with an intrinsic multi-photon component that scales with the pair generation rate \cite{Waks2006}. Furthermore, entanglement between the photon pairs limits indistinguishability \cite{Grice2001, Evans2010}. Correlated photon pairs can also be produced using four-wave mixing (FWM) in waveguides or fibers, although Raman scattering produces problematic background noise.

Single emitters, with no inherent capacity to produce multiple photons simultaneously, are more natural candidates for producing single-photon fields. Of these, atoms, ions and molecules  \cite{Kiraz2005} are appealing because the availability of identical systems guarantees the possibility of multiplexed indistinguishable photons \cite{Maunz2007, Eschner2009}. The outstanding challenge for each of these single-emitters is to efficiently capture emitted photons in a convenient spatial mode, although rapid progress is being made with high-finesse cavities and large numerical aperture optics \cite{Maiwald2012a, Leuchs2012a, Slodicka2015a}.

Atoms with $\Lambda$-type level schemes have been optically trapped inside high-finesse cavities for the production of photons by stimulated Raman adiabatic passage \cite{McKeever2003, McKeever2004a, Reiserer2014}. The single photon production probability in such a system is limited by the cavity mode density and optical cavity losses. Shallow optical traps further compromise neutral atom systems; the trap must be reloaded during operation and the possibility of capturing more than one atom contributes a multi-photon probability. In contrast, ion sources such as the one in this work can be held for longer periods and confined more closely in radio frequency (RF) traps. A charged particle source poses some challenges for strong cavity coupling, but both STIRAP and far off resonant Raman schemes have been performed with in-cavity generation efficiency close to 90\% \cite{Barros2009} .

Artificial atoms such as semiconductor quantum dots (QDs) \cite{Shields2007, Kako2006} can be integrated with optical microcavities \cite{Kress2005, Press2007, Fortsch2013a} to improve collection efficiency. While QDs are in principle single emitters, multi-photon noise in these systems is typically poor due to interactions with the bulk material \cite{Kuhlmann2013} and needs to be suppressed \cite{Somaschi2015}. Very recently QD sources have been engineered to produce consecutive photons with vanishingly low distinguishability \cite{Somaschi2015}, however scalable QD networks will require the capacity to tune each unique dot for producing mutually indistinguishable photons \cite{Patel2010}. Nitrogen vacancy (NV) centres in diamond nanocrystals are an alternative with similar distinguishability challenges. The single-photon purity of NV centre sources is currently obscured by the collection of scattered background light, but this is likely to improve with better coupling.

To compare the purity of single-photon states produced by various sources it is necessary to have a measure of quantum behaviour that captures the useful properties of the single photon field. The most ubiquitous measure of this non-classicality is the degree of anticorrelation $\alpha = P_c / P_s^2$ measured in a typical optical Hanbury Brown-Twiss (HBT) experiment, where $P_s$ and $P_c$ are the probability of single detection events and coincidence events respectively. For small mean photon flux the light field intensity autocorrelation $g^{(2)}(\tau)$ reduces to $\alpha$ when $\tau =0$ \cite{Kimble1977, Grangier1986}. However, this is just the first in a hierarchy of more stringent quantum signatures that can be used to test the performance of a single photon source \cite{Filip2013} and neglects the vacuum component of the optical state, which crucially characterizes the efficiency of photon generation by on-demand sources.

The negativity of the Wigner function is a stricter criterion for `non-classicality' and a necessary quantum computational resource \cite{Mari2012}. Wigner function negativity is a distinguishing property of all pure quantum non-Gaussian (QNG) states \cite{Lvovsky2001}, including ideal single-photon states. However, while a strongly attenuated field is necessarily non-negative in the Wigner representation, QNG remains an unambiguous and efficient test of higher-order quantum behavior even for mixed or attenuated states \cite{Straka2014}. Using the same HBT configuration as a typical $g^{(2)}(\tau)$ measurement \cite{Brown1956} one can measure photon detection probabilities sufficient for estimation of a QNG witness that distinguishes light fields from any convex mixture of coherent and squeezed states \cite{Filip2011}. Beating such a QNG threshold is therefore a necessary benchmark for applications in quantum information processing. This has stimulated several recent characterizations of single photon sources using the QNG measure for efficient but noisy sources \cite{Predojevic2012, Straka2014}.


\begin{figure}[t!]
\centerline{\includegraphics[width=\columnwidth]{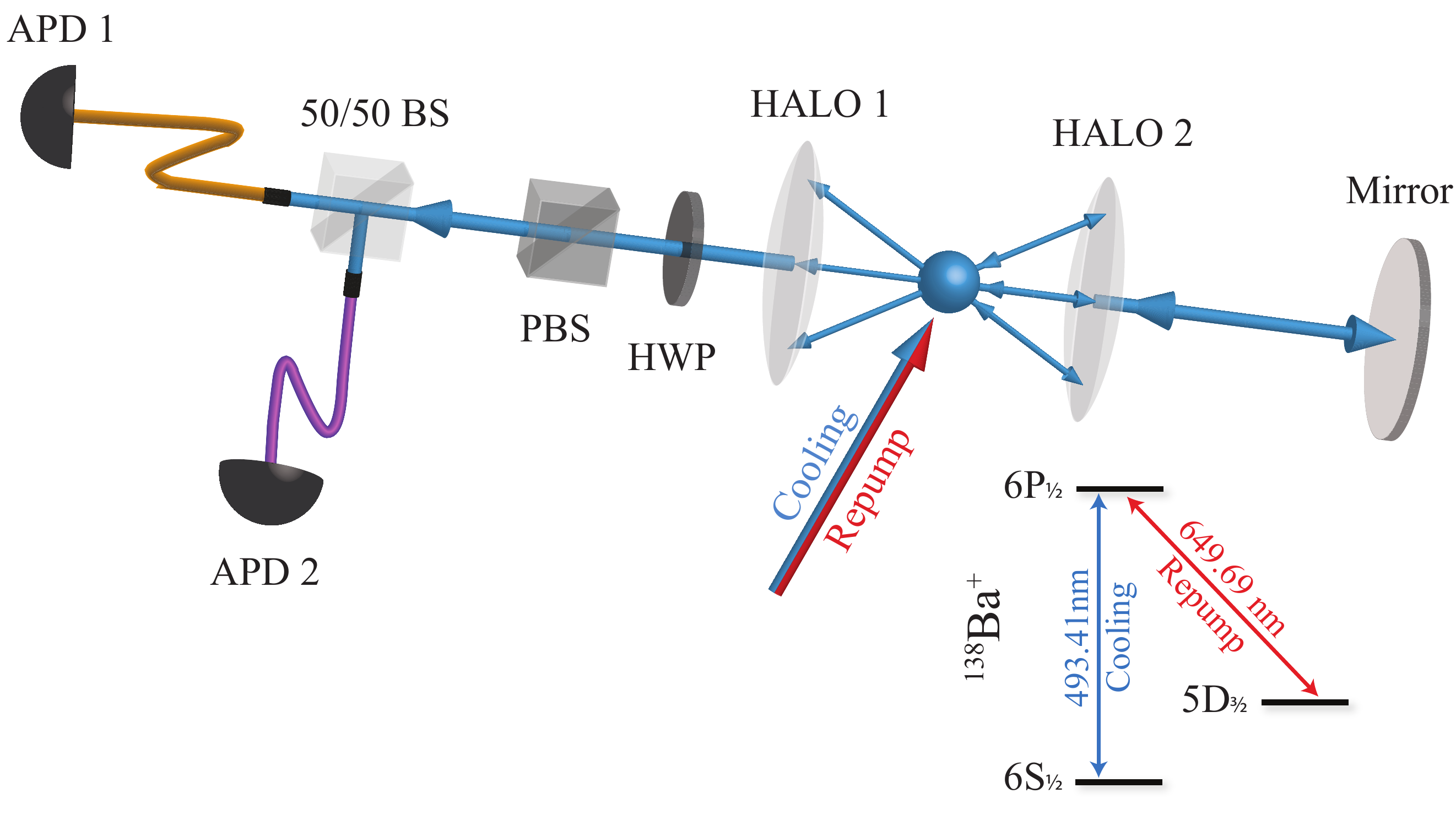}}
\caption{In this `reflected' configuration one side of the ion fluorescence is collimated by a high aperture lens objective (HALO) and reflected from a distant mirror. The reflected light is focussed back past the ion into the same spatial mode as fluorescence collected by a second HALO. The combined fields are polarization filtered and split between two fiber-coupled APDs. (inset) Electronic level structure of $^{138}$Ba$^+$.}
\label{experiment_fig}
\end{figure}

In this work we develop and characterize an experimental scheme for the generation of high purity on-demand single photons using a single trapped ion at the focus of high numerical aperture lenses under pulsed excitation. We apply both the conventional $g^{(2)}(0)$ and a new QNG metric to demonstrate that our source produces a temporally tunable single-photon light field. The source performs with near optimal QNG witness depth and is limited solely by the (extremely low) dark count rate of the detection technology available at this wavelength.

A $^{138}$Ba$^+$ ion is trapped and cooled in a linear Paul trap at the common focus of two high aperture lens objectives (HALOs) as per Fig.~\ref{experiment_fig}. The ion is Doppler cooled to well within the Lamb-Dicke regime using a 493 nm cooling laser and a repump beam at 650 nm to close the $6S_{1/2}$ - $6P_{1/2}$ - $5D_{3/2}$ cycle, Fig.~\ref{experiment_fig} (inset). Three pairs of Helmholtz coils create a magnetic field that defines the atomic quantization axis of the radiating dipole at $90^\circ$ to the common axis of the cooling and repump beams and lifts the degeneracy of the three atomic levels in the simple $\Lambda$ scheme in Fig.~\ref{experiment_fig} such that the complete system dynamics is described by a set of eight-level Bloch equations. Each HALO has a numerical aperture of 0.4 and together they collect a combined 12$\%$ of the ion's dipole fluorescence on the 493 nm Doppler cooling transition.


The configuration is a traditional Hanbury Brown-Twiss (HBT)
measurement with one reflected field; two low-noise fiber-coupled
APDs with quantum efficiencies of 70\%  and 73\% on either arm of
a 50/50 beamsplitter sample a single spatial and polarization mode
of the ion fluorescence collected by one of the HALOs.
Fluorescence collected by the second HALO is back-reflected over
the ion by a distant mirror such that it too is collected by the
same APDs. The ion fluorescence in this configuration self
interferes with 90\% visibility when sideband cooled
\cite{Slodicka2012a} and 70\% visibility when Doppler cooled as it
is here \cite{Eschner2001}. Locking the ion-mirror path length for
constructive interference enhances the detection efficiency, but
the reflected configuration results presented here are recorded
with the mirror position freely scanning such that
self-interference effects average out. Detection events at the
APDs are time tagged using a Picoharp 300 photon counting module
with a time resolution of 4 ps.

\begin{figure}[t!]
\centerline{\includegraphics[width=\columnwidth]{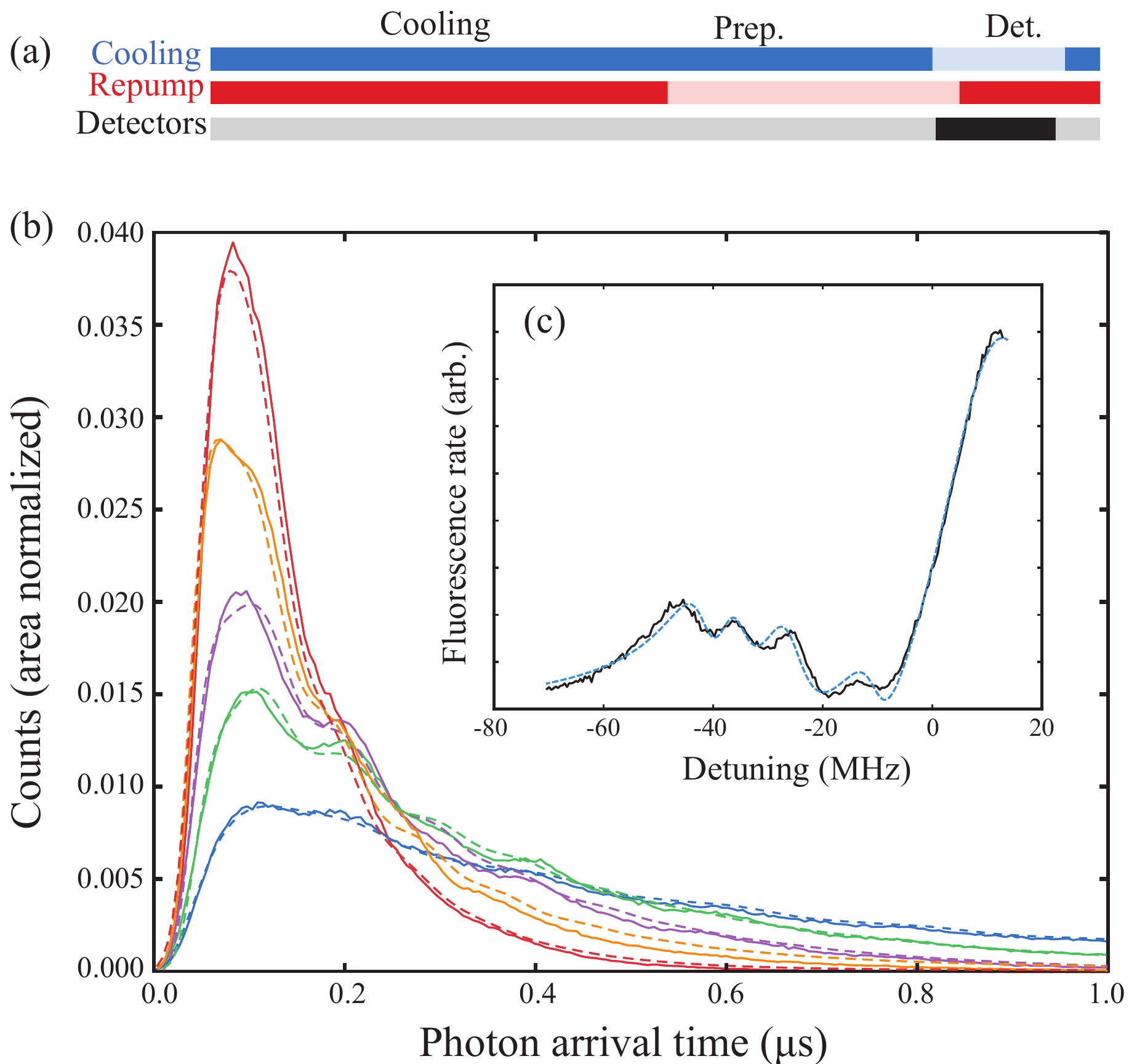}}
\caption{(a) The Raman driving scheme that is used to cool, prepare and trigger the trapped ion photon source. Single photons are produced on demand by alternating the cooling (blue) and repump (red) beams. Both beams are on for most of the cycle to Doppler cool the ion. During the preparation stage the repump beam is switched and the ion's electron is shelved in a D$_{3/2}$ dark state. A single blue photon is emitted using only the repump beam and detected by fiber coupled APDs. (b) Area normalised histograms of single-photon arrival times produced using the above scheme. These photon intensity distributions show the mean arrival time decreasing as the repump beam power $\Omega_r$ is steadily increased to the transition saturation point. The eight-level Bloch model of our system predicts corresponding single photon shapes shown by dashed lines. For each trace the model parameters are calibrated using a dark resonance scan under the same driving conditions. A typical example of the dark resonance parameter fit is shown in inset (c) with the measured fluorescence rate (solid black line) and fit from the model (dashed blue line) with parameters corresponding to the slowest (blue) single photon shape.}
\label{shape_fig}
\end{figure}

Photons are produced on demand at a rate of 200 kHz by preparing the ion in a mixture of metastable $5$D$_{3/2}$ Zeeman states and then triggering photon emission from the 493 nm cooling transition by alternating the cooling and repump beams according to the scheme shown in Fig.~\ref{shape_fig}a. The experimental sequence begins with 2 $\mu$s of Doppler cooling with both the cooling and repump beams, then 1 $\mu$s of optical pumping with only the cooling beam, preparing the ion in a D$_{3/2}$ shelving state. Both beams are off for 500 ns before the repump beam is switched back on (rise time 90 ns) to trigger the emission of a 493 nm photon. In Fig.~\ref{shape_fig}b we show arrival time histograms of photons produced in this way as the repump power is steadily increased to the transition saturation point. Increasing repump power shortens the pulse length with a corresponding reduction in dark counts during the detection window.

The magnetic field magnitude ($\vec{B}$) is calibrated by spectroscopy of the quadrupole transition 6S$_{1/2}$ $\rightarrow$ 5D$_{3/2}$ using a fiber-coupled 1.7$\mu$m laser locked to a narrow linewidth cavity. The cooling and repump beam powers ($\Omega_g$, $\Omega_r$), the detuning ($\Delta_g$) and common polarisation ($\vec{P}$) are inferred from a spectroscopic dark state measurement prior to operation. Complete eight-level Bloch simulations predict the stationary fluorescence rate as a function of $\Delta_r$ and the above parameters are fitted to the corresponding measurement, see Fig.~\ref{shape_fig}c. The same Bloch equations can then be used to calculate the dynamic single-photon scattering rate as a function of time after the repump is switched on under the driving scheme in Fig.~\ref{shape_fig}a, with repump beam parameters ($\Omega_r$, $\Delta_r$, $\vec{P}$) inferred from the continuous-driving dark state scan. This dynamic simulation takes into account the intensity profile of the repump beam (rise time 90 ns) as measured after the switching/shaping AOM, and assumes that the preparation process leaves the ion in a mixed state with probabilities equally distributed between the four $5D_{3/2}$ levels and with no coherences between that manifold. Each of these initial D states produces photons with similar wavepackets such that consecutive photons have negligible distinguishability \cite{Eschner2009}.

Taking into account the detection mode polarisation, which has been optimised to maximise detection probability, the eight-level optical Bloch model predicts dynamic scattering probabilities in close agreement with the measured single photon wavepacket (Fig.~\ref{shape_fig}a). The arrival time distribution features a distinctive oscillation at a frequency independent of the repump beam power. These quantum beats are caused by interference between $6P_{1/2}$ - $5D_{3/2}$ absorption amplitudes that enhances and suppresses the emission of Raman-scattered $6S_{1/2}$ - $6P_{1/2}$ photons in the detection mode with frequency determined by the D state energy splitting \cite{Schug2014}.

\begin{figure}[t] \centerline{\includegraphics[width=\columnwidth]{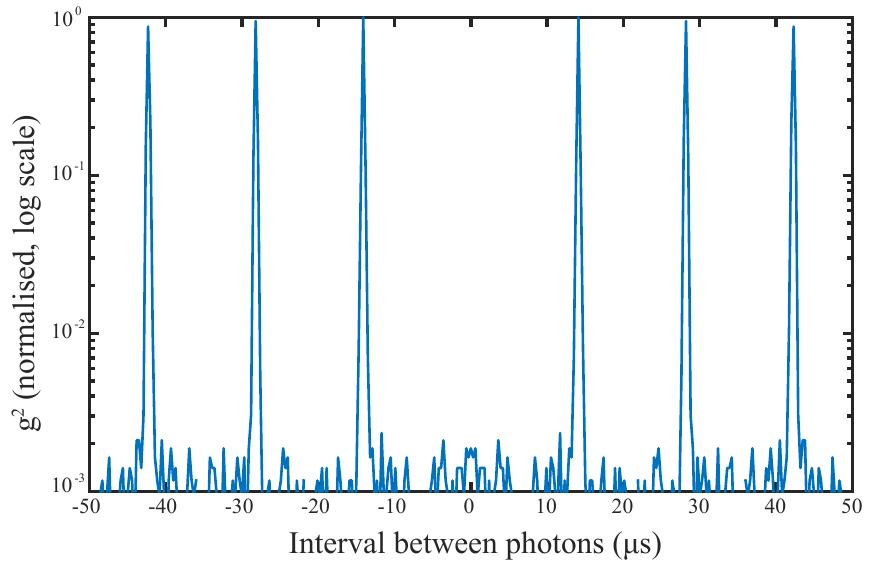}} \caption{Intensity auto-correlation ($g^2 (\tau)$) of the pulsed resonance fluorescence when the trapped-ion single photon source is operated at a rate of 70 kHz. The data is presented without background subtraction and shows g$^2(0) = (1.9 \pm 0.2) \times 10^{-3}$. The pulsed source is triggered according to the scheme in (Fig.~\ref{shape_fig}b) but at one third of the optimal repetition rate so that a statistically significant amount of background can be measured during an extended detection window. Fluorescence in the cooling periods between the photon generation windows has been removed.} \label{g2_fig} \end{figure}

The second order intensity correlation function $g^2 (0) = 0.0019(2)$ of the field emitted by our photon source violates the coherent state condition $\alpha \leq1$ under continuous driving, and also when pulsed with the above scheme to produce single photons on demand (Fig.~\ref{g2_fig}). The measured $\alpha$ corresponds precisely to the stated and measured dark count rate of our two detectors, $10\pm2$ counts/s per detector. By subtracting this measured dark count rate from the measured coincidences we can say
with $95\%$ confidence that the intrinsic $\alpha$ of our source is below $3\times10^{-4}$. Such photon purity bodes well for the scalability of networks based on atomic sources.

\begin{figure}[t]
\centerline{\includegraphics[width=\columnwidth]{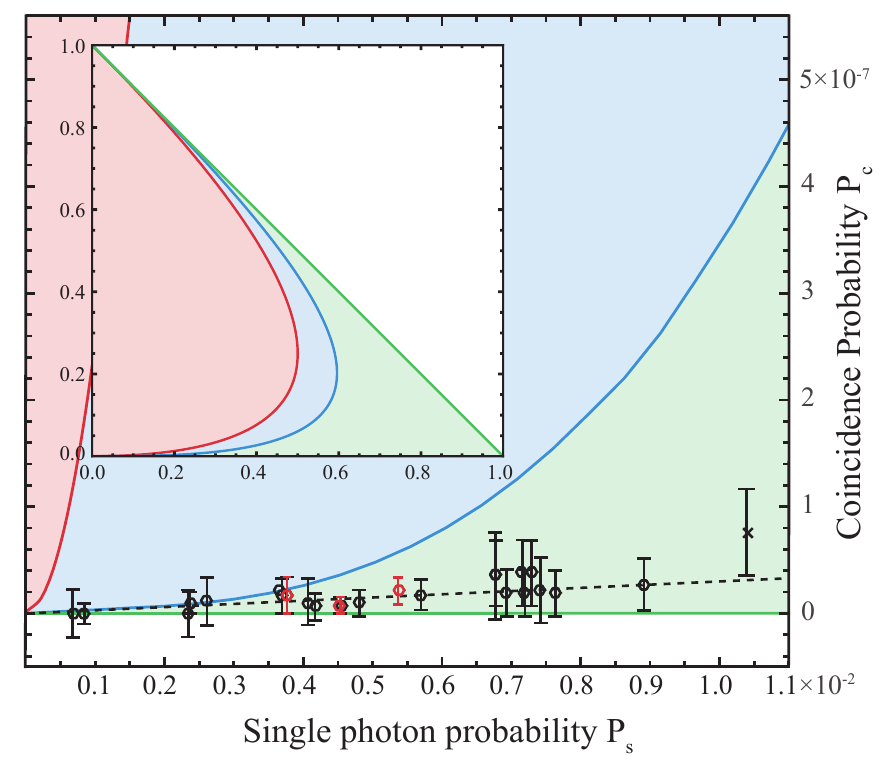}}
\caption{Coincidence probability $P_c$ versus single photon detection probability $P_s$ in the operational region of our source and over the complete phase space (inset). Classical light coincidence rates cannot be reduced below the NC threshold (red) and Gaussian light fields are similarly restricted to coincidence rates above the QNG threshold (blue). Light fields with coincidence rates below this threshold (green region) are unambiguously QNG. Circles correspond to measurements of our pulsed single-photon source in the reflected (red) and symmetric (black) configuration with a 200 ns detection window and under a range of driving conditions. The black cross is a measurement under optimal conditions with a 500 ns detection window. Error bars represent 95\% confidence intervals. The dashed line predicts the performance of an ideal single-photon source measured with our detectors and a 200 ns detection window.}
\label{dw_fig}
\end{figure}

We further analyze the quality of the single photons
produced by our source by measuring a signature of its
higher order quantum behaviour, the QNG state witness
\cite{Filip2011}. Following Ref. \cite{Lachman2013}, both the
traditional non-classicality (NC) criteria $\alpha < 1$ and this
QNG witness can be formulated in terms of probabilities $P_s$ and $P_c$ measured in an optical HBT configuration. Classical states (states that can be written as a mixture of coherent states) are bounded
by the inequality $P_s \leq 2(\sqrt{P_c} - P_c)$ illustrated by
the red curve in Fig.~\ref{dw_fig} (inset). A stricter threshold
can be calculated for Gaussian states \cite{Lachman2013}, blue
curve in Fig.~\ref{dw_fig}(inset). The QNG threshold can be
written for these observables in an implicit form parametrized by the degree of squeezing 
(V) of the threshold states (derivation presented in supplementary material)



\begin{align}
P_s &= \frac{1}{2} + \frac{(1-V(2+V)) e^{\frac{V-1}{2V}}} {\sqrt{V}(1+V)^2}  \\
P_c &=\frac{1}{2} - \frac{(1+V^2) e^{ \frac{V-1}{2V}}}{\sqrt{V}(1+V)^2}
\end{align}

In the operational region of our source this threshold can be reduced to $P_c \approx P_s^3/3$, see Fig.~\ref{dw_fig}. Any state with single-mode coincidence probability $P_c$ below this threshold is unambiguously QNG. Our photon source is strongly attenuated by the collection efficiency of the HALOs, fiber coupling mode mismatch and detection efficiency. For this reason the vacuum term dominates in a Fock basis description of our state and our source sits in a corner of the $P_s$, $P_c$ phase space. Nonetheless, because our source is pure it beats the QNG threshold by 6 s.d. in the optimal configuration, Fig.~\ref{dw_fig}(red circles). The number of detections and coincidences are determined in a detection window 200 ns long from the beginning of the photon trigger. These measurements further confirm an excellent quality of the generated single photons, limited solely by the overall single-photon collection efficiency and detector dark counts.



\begin{figure}[t!]
\centerline{\includegraphics[width=\columnwidth]{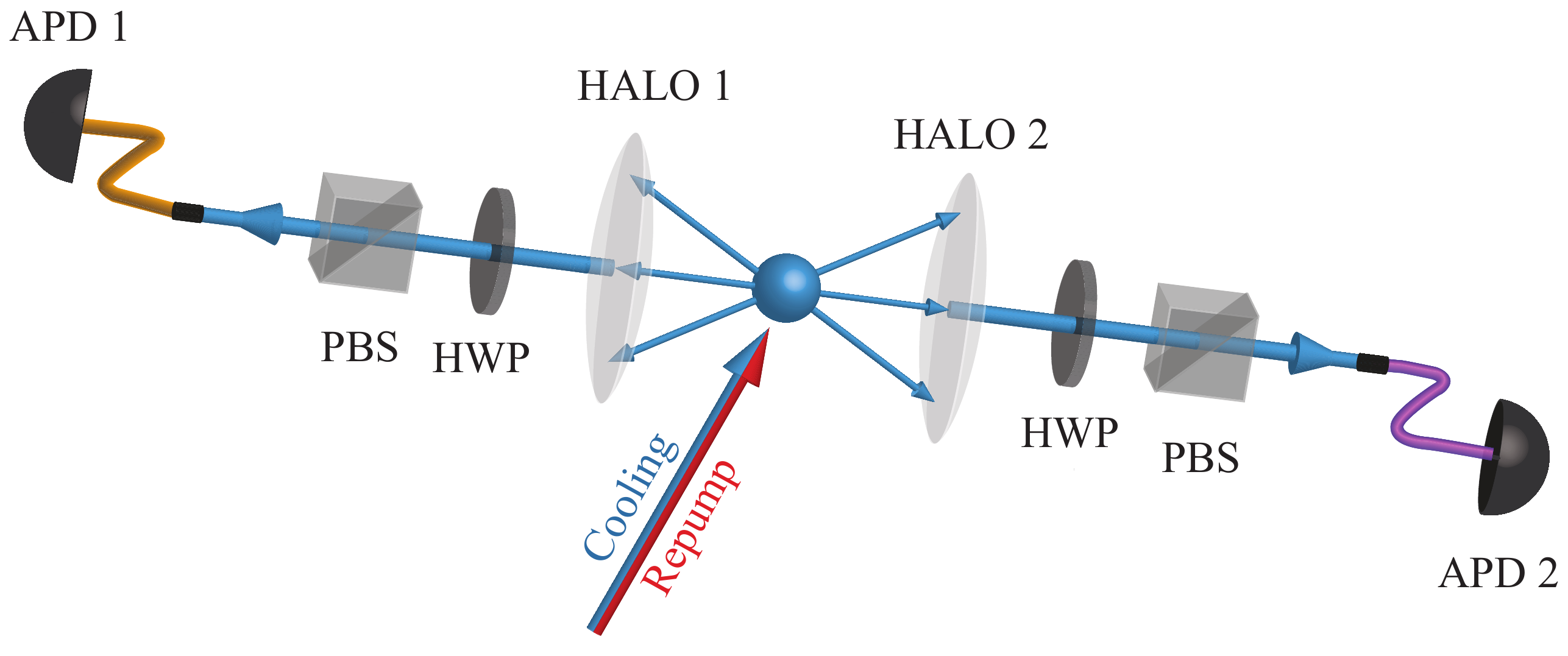}}
\caption{In the `symmetric' configuration fiber-coupled APDs sample separated spatial and polarization modes from opposite sides of the ion simultaneously.}
\label{sym_fig}
\end{figure}

The typical approach to enhancing collection efficiency from low-efficiency single photon emitters employs combinations of high numerical aperture optical elements \cite{Maiwald2012, Gerber2009a, Kurz2014, Shu2010, Streed2011a}. This corresponds to simultaneous emission of a light field into several spatial modes. Such a multi-mode source is useful so long as the purity of emitted single-photons is not compromised. Simultaneous enhancement of spurious background light collection, spatial restrictions on excitation beams, excitation beam scattering into the photon collection modes and other processes can in general make enhancing collection efficiency unfavorable for the photon purity. To estimate this effect, we measure the same QNG witness for a light state emitted coherently by a single-photon source into two spatial modes. We apply this measure to our single trapped ion source as presented in Fig.~\ref{sym_fig}. The fluorescence is detected in two spatial directions by APDs positioned behind each HALO lens. In the reflected configuration, the collected photons were forced to enter the same spatial mode prior to the detection, in
the symmetric setup they are emitted into two, in principle independent, directions. Such an arrangement also reduces absorption losses from the extra optical elements and increases coupling efficiency compared to the reflected configuration. As demonstrated in \cite{Slodicka2012a,Eschner2001}, ion fluorescence emitted in two opposite directions is mutually phase coherent to a very high degree and can be efficiently transferred into a single spatial mode, for example by recombination on a beamsplitter. The single ion in the symmetric configuration thus comprises both the emitter and perfect unitary beamsplitter of the typical Hanbury Brown-Twiss measurement configuration. The QNG measurements presented for our source emitting into a single spatial mode can be compared to two-mode coincidence measurement evaluated to yield the same measure.

Measurements made in the symmetric configuration with a detection window of 200 ns, Fig.~\ref{dw_fig}(black circles) suggest that the fluorescence collected in the two modes still corresponds to extremely pure single photons, more than 20 s.d. from the QNG threshold. Increasing the detection window to 500 ns, Fig.~\ref{dw_fig}(black x), improves the proportion of photons detected with an increase in the coincidence rate proportional to the detection time. These measurements indicate that the techniques currently being pursued to increase collection efficiency from single emitters in free space won't have any detrimental effect on the purity of the collected photons.


In both the symmetric and reflected configurations, and for a wide range of driving
conditions, the trapped-ion photon source produces a pulsed light field with coincidence rates indistinguishable from an ideal source measured with our detectors, see Fig.~\ref{dw_fig}(dashed
line). The presented quality of single-atom generated single photons, together with recent technological efforts to enhance the collection efficiency of single-atom fluorescence\cite{Maiwald2012a, Fischer2014, Wong-Campos2015, Streed2012} foretells decisive progress in overcoming one of the greatest remaining obstacles for practical scalability of photonic quantum information technology, the lack of on-demand pure single photons\cite{Eisaman2011}. Our demonstration shows the first unambiguous proof that very pure single photons with no intrinsic two-photon contributions can be generated using conceptually simple and well developed technology. Improved collection from such a source will facilitate completion of a long-standing pursuit of the broad quantum optics community, the efficient generation of single photons.

\vspace{5mm}

\section*{ACKNOWLEDGMENTS}
The work reported here has been supported by the Austrian Science Fund FWF (SINFONIA, SFB FoQuS), by the European Union (CRYTERION \#227959), and by the Institut f\"{u}r Quanteninformation GmbH.

\bibliographystyle{unsrt}
\bibliography{library}

\clearpage

\section{Supplementary material: Derivation of QNG threshold}

The quantum non-Gaussian (QNG) witness presented in this paper differs in form from similar thresholds presented before \cite{ Filip2011, Lachman2013}. The witness distinguishes optical QNG states by the probability $P_c$ of coincident photon detections, conditioned on the probability $P_s$ of single photon detection in an optical HBT. The derivation of this threshold follows the method of \cite{Lachman2013}, with the threshold formulated in terms of observables directly measured in this experiment.

The Wigner function of the vacuum state is

\begin{equation}
 W_{\mbox{vac}}(x,p)=\frac{1}{2\pi}e^{-\frac{x^2+p^2}{2}},
 \end{equation}
 where $x$ and $p$ are generalized coordinates for position and momentum. The Wigner function of a general, pure squeezed state is reached by squeezing the variance of the x and p coordinates by parameter V
 \begin{equation}
 W_V(x,p)= W_{\mbox{vac}}\left(\frac{x}{\sqrt{V}},\sqrt{V}p\right),
 \end{equation}
followed by rotation by an angle $\phi$
 \begin{equation}
 W_{V,\phi}(x,p)=W_V(x \cos \phi + p \sin \phi,- x \sin \phi +p \cos \phi)
 \end{equation}
and displacement in $x$ coordinate by distance $r$
\begin{equation}
W_{V,\phi,r}=W_{V,\phi}(x-\sqrt{r},p).
\end{equation}

The probability $P_n$ of measuring this general, pure squeezed state with photon number $n$ is the overlap of the state with projectors $W_n$ for the $n$ photon Fock state. The projectors of the zero and one photon Fock states are \cite{Hillery1984}

\begin{eqnarray}
W_0(x,p)&=&2 e^{-\frac{x^2+p^2}{2}}\nonumber \\
W_1(x,p)&=&2 (x^2+p^2-1) e^{-\frac{x^2+p^2}{2}}.
\end{eqnarray}

Integration $P_{n}=\int_{-\infty}^{\infty} W_{n}W_{V,\phi,r} \mathrm{d}x \mathrm{d}p$ yields the vacuum, single photon and higher order Fock state probabilities \cite{Yuen1976}

\begin{align}\label{eq:Ps}
& P_0 = \frac{2\sqrt{V}}{1+V} e^{\frac{r(-1-V + (V-1)\cos{\left(2\phi\right)}}{4(1+V)}} \\
& P_1 = \frac{r(1+V^2 - (V^2-1) \cos{(2 \phi}))}{2(1+V)^2}P_0 \\
& P_{2+}  = 1 - P_0 - P_1
\end{align} 

Taking into account the probability of false single photon measurements, in which two photons arrive simultaneously at the same detector, the measurement probabilities $P_s$ and $P_c$ are related to the Fock state probabilities by 

\begin{align}
P_s &= P_1 + (P_{2+}/2)\\
P_c &= P_{2+}/2
\end{align} 

To derive the QNG threshold for these measured experimental probabilities we consider the linear functional

\begin{align}
F(a) &= P_s + aP_c
\end{align} 

where a is a free parameter. Following the method of \cite{Lachman2013} we see that the squeezed states that maximise this linear, unconstrained function are the border states forming the QNG threshold in the directly measurable $P_c$, $P_s$ parameter space. For a given $P_c$ these states have the maximum $P_s$ possible on the set of Gaussian states. The angle $\phi$ between the squeezing and displacement directions appears only in the argument of the cosine, so $F(a)$ is necessarily maximised by $\phi = 0$. The maximum of the functional also satisfies $\frac{\partial F}{\partial V} = 0$ and $\frac{\partial F}{\partial r} = 0$ which gives

\begin{align}
\frac{\partial P_s}{\partial V} & = - a \frac{\partial P_c}{\partial V} \\
-a \frac{\partial P_c}{\partial r} & = \frac{\partial P_s}{\partial r}.
\end{align}

Together these give the condition

\begin{align}
\frac{\partial P_s}{\partial V} \frac{\partial P_c}{\partial r}&= \frac{\partial P_s}{\partial r}\frac{\partial P_e}{\partial V}.
\end{align} 

For $P_s$ and $P_c$ of our general squeezed state this condition evaluates to

\begin{equation}
\frac{V^2+rV-1}{(1+V)^6}e^{-\frac{r}{1+V}} = 0
\end{equation}

which is satisfied only when

\begin{equation}
r = \frac{1-V^2}{V}.
\end{equation} 

with this $r$-$V$ relation the threshold $P_c$, $P_s$ values can be reduced to the pair of equations presented in the main body of the paper.

\begin{align}
P_s &= \frac{1}{2} + \frac{(1-V(2+V)) e^{\frac{V-1}{2V}}} {\sqrt{V}(1+V)^2}  \\
P_c &=\frac{1}{2} - \frac{(1+V^2) e^{ \frac{V-1}{2V}}}{\sqrt{V}(1+V)^2}
\end{align}

\end{document}